\documentstyle[preprint,aps]{revtex}
\begin{document}
\begin{titlepage}
\begin{center}
\today     \hfill    MIT-CTP-2369 \\
           \hfill    hep-ph/9412322 \\
\vskip .5in
{\large {\bf Naturalness and superpartner masses\\ or  \\
 When to give up on weak scale supersymmetry  }}
\vskip .5in
Greg W. Anderson \footnote {email address: anderson@fnth03.fnal.gov}
and Diego J. Casta\~no
\footnote {email address: castano@fshewj.hep.fsu.edu}\\
{\em Center for Theoretical Physics\\
Laboratory for Nuclear Science \\
 Massachusetts Inst. of Technology\\
Cambridge, MA 02139.}\\
\end{center}

\begin{center}
{\it To appear in:} Phys. Rev. {\bf D52}, 1693 (1995)
\end{center}
\vskip .5in

\begin{abstract}
Superpartner masses cannot be arbitrarily heavy if supersymmetric
extensions of the standard model explain the stability of the gauge
hierarchy.  This ancient and hallowed motivation for weak scale
supersymmetry is often quoted, yet no reliable determination of this
upper limit on superpartner masses exists.  In this paper we compute
upper bounds on superpartner masses in the minimal supersymmetric model,
and we identify which values of the superpartner masses correspond
to the most natural explanation of the hierarchy stability.  We compare
the most natural value of these masses and their upper limits to the
physics reach of current and future colliders.  As a result, we find that
supersymmetry could explain weak scale stability naturally even if no
superpartners are discovered at LEP II or the Tevatron (even with the
Main Injector upgrade).  However, we find that supersymmetry cannot
provide a complete explanation of weak scale stability, if
squarks and gluinos have masses beyond the physics
reach of the LHC.  Moreover, in the most natural scenarios, many sparticles,
for example, charginos, squarks, and gluinos, lie within the physics reach
of either LEP II or the Tevatron.  Our analysis determines the most natural
value of the chargino (squark) ((gluino)) mass consistent with current
experimental constraints is $\sim$ 50 (250) ((250))  GeV and the corresponding
theoretical upper bound is $\sim$ 250 (700) ((800)) GeV.
\end{abstract}
\end{titlepage}
\newpage
\mbox{ }
\setcounter{page}{1}

\section{Introduction}
\indent

As a candidate for physics beyond the standard model,
weak scale supersymmetry has several appealing features:
It provides an understanding of why a light weak scale is stable,
it successfully predicts the value of $\sin^2\theta_W$
assuming gauge unification, it predicts a top quark Yukawa
coupling of order one, leading to a heavy $M_t$ (assuming $\tau$ lepton
and bottom quark Yukawa coupling unification), and it provides a natural
cold dark matter candidate in the form of the lightest superpartner.

Despite these circumstantial arguments for weak scale supersymmetry, there
is not a shred of {\it direct} experimental evidence to support it.
Should we be surprised or discouraged that we have not yet
found any supersymmetric partners to the standard model
particles?   To date, of the particles we believe to be fundamental,
all those observed would be massless if the gauge symmetries of the standard
model were unbroken.~\footnote{Only quite recently has there
been experimental evidence
for the top quark.}  Because the current,
experimental probes only reach up through
the lower fringes of the weak scale, it is not surprising that the
fundamental particles discovered so far obtain masses as a consequence of
spontaneously broken gauge symmetries.  Their superpartners, by contrast,
can have gauge invariant mass terms, provided supersymmetry is broken.
Although they are not required to be light by gauge
symmetries, there is a theoretical upper limit on their masses
above which the weak scale does not arise naturally.
As the scale of supersymmetry breaking is increased,
the weak scale can only remain light by virtue of an
increasingly delicate cancellation.  Requiring that the weak scale
arises naturally places an upper bound on superpartner masses.

In this paper, we attempt to quantify the relationship
between naturalness and superpartner masses.
Using recently formulated naturalness measures
we compute the most natural value of the superpartner masses,
the extent to which naturalness is lost as experimental bounds
on superpartner masses increase, and a theoretical upper limit
to the masses of superparticles.

In section two we review the
naturalness measures used in our study.
Section three is devoted to a discussion of radiative electroweak
symmetry breaking in the minimal supersymmetric extension of the Standard
Model (MSSM) and to details of the numerical methods we
employed in our analysis.  The results, presented in section four,
demonstrate that the MSSM can not
accommodate the weak scale naturally if superpartner masses lie beyond
the reach of LHC.  Moreover, in the most natural cases,
physics beyond the standard model has a good chance of being discovered
at LEP II or the Tevatron.
\section{Measuring Fine Tuning }
\indent
  In this section we review the recently formulated naturalness measures
we use in our analysis.  A more detailed motivation and derivation of
these criteria can be found in Ref. [1].  Any measure of naturalness
contains  assumptions about how the fundamental parameters of a
Lagrangian are distributed.  If we parameterize these assumptions,
a quantitative
measure of naturalness follows directly.  Consider a Lagrangian
density written in terms of fundamental couplings specified
at the high energy boundary of the effective theory:
${\cal L}(a_1,a_2,...a_n)$.  At a low energy scale, we can write the
Lagrangian in terms of physical observables $X$ ({\it e.g.}, $X=M_{Z}^2$).
These observables will depend on the $a_i$ through the renormalization
group and possibly on a set of minimization conditions: $X = X(a)$.
If we assume the probability distribution of a fundamental
Lagrangian parameter $a$ is given by
\begin{eqnarray}
dP(a) = \frac{f(a) da}{\int f(a) da},
\eqnum{2.1}
\end{eqnarray}
a likelihood distribution for the low energy observable $X$ follows
\begin{eqnarray}
\int_{a_{-}}^{a_{+}} f(a) da = \int_{X(a_{-})}^{X(a_{+})}\rho(X)\,dX \ .
\eqnum{2.2}
\end{eqnarray}
The  value of an observable $X$ is unnatural if it is relatively
unlikely to end up in an interval $u(X)$ about $X$
compared to similarly defined intervals around other values of $X$.
The probability that $X$ lies within an interval $u(X)$ about
$X$ has weight $u \rho$.  So
we define our quantitative measure of naturalness as
\begin{eqnarray}
\gamma = { \langle u\rho \rangle \over u(X)\rho(X) },
\eqnum{2.3}
\end{eqnarray}
where
\begin{eqnarray}
\langle u \rho \rangle = {\int u\rho \, da \over \int da }.
\eqnum{2.4}
\end{eqnarray}
The conventional sense of naturalness
for hierarchy problems corresponds to an interval
$u = X$.~\footnote{
For example, in a theory of fundamental scalars, the scalar
mass is related to the cut-off $\Lambda$ and the bare term
$m_0$ by: $m_{s}^2 = g^2 \Lambda^{2} - m_{0}^2$. In this
theory we must adjust $g^2$ with the same precision to place
the scalar mass squared in a $1\, {\rm (GeV) } ^{2}$ window
whether the scalar mass is $O(\Lambda)$ or $O(10^{-14}\Lambda)$.
A small mass for the scalar is unnatural in the sense that a
small change in $g^2$ leads to a large {\it fractional}
change in $m_{s}^2$ ~\cite{AC1}. }
With this prescription, fine tuning corresponds to $\gamma >> 1$.
The $\gamma$ defined in Eq. (2.3) is proportional to the
Barbieri-Giudice sensitivity
parameter $c(X,a) = |(a/X)(\partial X /\partial a)|$ \cite{BG}.  We
can use Eqs. (2.3-4) to define an average sensitivity $\bar{c}$ through
the relation
\begin{eqnarray}
\gamma = c/\bar{c} \ .
\eqnum{2.5}
\end{eqnarray}
This definition of $\bar{c}$ gives
\begin{eqnarray}
1 /\bar{c}  = \frac{\int da \, a f(a) \, c(X;a)^{-1}}
{a f(a)\int da}.
\eqnum{2.6}
\end{eqnarray}
The naturalness measures defined by Eqs. (2.5-6) are a
refinement of Susskind's description of Wilson's
naturalness criteria \cite{Wilson}:
Observable properties of a system,
{\it i.e.,} $X$, should not be unusually unstable with respect
to minute variations in the fundamental parameters, $a$.
In other words, $X(a)$ is fine tuned if the values of the
fundamental parameters $a$ are chosen so that $X$ depends
on the $a$ in an {\it unusually} sensitive manner when compared to
other values of the fundamental parameters $a$.  Sensitivity in
this case is understood to mean that a small {\it fractional} change in $a$
leads to a large {\it fractional} change in $X$.~\footnote{
In deriving the naturalness criteria Eqs (2.3-2.6), we have
attempted to make explicit the discretionary choices
inherent in any quantitative measure of naturalness.  In
any particular application, in order to obtain a reliable measure
of naturalness, these choices must be made sensibly. }

Returning to Eq. (2.4-2.6), we see that
three choices need to be specified
before we can make practical use of this prescription.
First, the choice of $f(a)$ reflects our theoretical
prejudice about what constitutes a natural value of the Lagrangian
parameter $a$.  We will make two different choices for
$f(a)$ as an aid in determining how sensitively the bounds we
derive depend on this theoretical prejudice: $f(a) = 1$ and
$f(a)=1/a$.  We denote the corresponding naturalness measures
by $\gamma_1$ and $\gamma_2$, respectively.  The bounds we
derive on superpartner masses in section four are fairly insensitive
to this choice.
Second, the conventional
notion of naturalness for hierarchy problems is
$u(X) = X$~\cite{AC1}.  This choice has already been made in Eq.
(2.6) and it is implicit in the qualitative statement of naturalness
written above.   Finally, the range of integration $(a_{-},a_{+})$
for the averaging must be chosen.  This range will be discussed
in section four.

\section{The MSSM }
\indent All the chiral interactions of the MSSM
are described by its superpotential
\begin{eqnarray}
   W = {\hat{\overline u}} {\bf Y}_u {\hat \Phi}_u {\hat Q} +
       {\hat{\overline d}} {\bf Y}_d {\hat \Phi}_d {\hat Q} +
       {\hat{\overline e}} {\bf Y}_e {\hat \Phi}_d {\hat L} +
       \mu {\hat\Phi}_u {\hat\Phi}_d \ .
\label{superpotential}
\eqnum{3.1}
\end{eqnarray}
The $\mu$-term explicitly breaks the Peccei-Quinn symmetry and avoids
a phenomenologically disastrous axion.  In addition to all the particles
of the SM, there are thirty-one new ones including three new Higgs
bosons.

Supersymmetry is explicitly broken in the MSSM using soft terms derived
from the low energy limit of supergravity (SUGRA) theory.  The form of
the soft SUSY breaking potential in the MSSM includes mass terms for all
the scalars and for the gauginos as well as bilinear and trilinear terms
following from the K\"ahler potential of the SUGRA theory in the low
energy limit.

A generic feature of minimal low energy SUGRA models is universality of
the soft terms.  Universality implies that all the scalar mass
parameters are equal to the gravitino mass, $m_0$, at some high
energy scale which we take to be the scale of gauge coupling unification,
$M_X=10^{16}$ GeV.  All soft trilinear couplings share a common value,
$A_0$, that can be related to the soft bilinear coupling, $B_0$,
depending on the form of the K\"ahler potential.  To some extent,
universality in the soft breaking terms is required in order to avoid
unwanted flavor changing neutral current effects.  Since the gauge
couplings unify, the gaugino mass parameters are assumed equal to a common
value, $m_{1/2}$, at $M_X$.  Consequently, the minimal model introduces
five new parameters, $m_0$, $A_0$, $m_{1/2}$, $B_0$, and $\mu_0$.  However,
it is very predictive since these account for the masses of thirty-one new
particles \cite{mr}.

In the MSSM, the electroweak symmetry is broken
radiatively \cite{ir,ikkt,agpw,ehnt}.
In our analysis, we use the 1-loop effective Higgs potential
\begin{eqnarray}
   V_{1-{\rm loop}} = V_0 + \Delta V_1 \ ,
\label{vhiggs}
\eqnum{3.2}
\end{eqnarray}
where the expression for the 1-loop correction is given by
\begin{eqnarray}
   \Delta V_1 = {1\over64\pi^2} \sum_{i} (-1)^{2s_i} ( 2s_i + 1 )
                m_i^4 ( {\rm ln}{m_i^2\over Q^2} - {3\over2} ) \ .
\label{dv1}
\eqnum{2.3}
\end{eqnarray}
The $m_i$ represent the field dependent masses of the particles
of the model and the $s_i$ the associated spins.  We include the
contributions of all the MSSM particles in the 1-loop correction.

Using the renormalization group, the parameters are evolved to low energies
where the potential attains validity.  This RG improvement uncovers
electroweak symmetry breaking.  The exact low energy scale at which to
minimize is unimportant as long as the 1-loop effective potential is used
and the scale is in the expected electroweak range.  The minimization scale
will arbitrarily be taken to be $M_Z$.  If the electroweak symmetry is broken,
minimization yields non-zero values for the vacuum expectation values (VEVs)
of the two Higgs fields, $v_u$ and $v_d$, or equivalently
$v=\sqrt{v_u^2+v_d^2}$ and $\tan\beta=v_u/v_d$.
The two minimization conditions may be expressed as follows
\begin{eqnarray}
   \mu^2(M_Z) &=& {{\overline m}_{\Phi_d}^2 -
   {\overline m}_{\Phi_u}^2 \tan^2\beta \over \tan^2\beta - 1 }
   - {1\over2} m_Z^2 \ , \label{min1}\eqnum{3.4} \\
   B(M_Z) &=& { ({\overline m}_{\Phi_u}^2 + {\overline m}_{\Phi_d}^2
              + 2\mu^2)\sin2\beta \over
              2 \mu(M_Z) } \ , \label{min2}
\eqnum{3.5}
\end{eqnarray}
where ${\overline m}_{\Phi_{u,d}}^2 = m_{\Phi_{u,d}}^2 +\partial\Delta
V_1/\partial v_{u,d}^2$.  In the process of integrating the 2-loop
renormalization group equations, the threshold corrections
due to all the light particles are implemented as step functions \cite{cpr}.

The procedure we follow to analyze the MSSM assumes the following
4+1 free input parameters:
$A_0$, $m_0$, $m_{1/2}$, $\tan\beta_0$, and ${\rm sign}(\mu)$ since it is
undetermined from Eq.~(3.4).
The other parameters of the MSSM are fixed using the following constraints:
Electroweak breaking in the form of two minimization conditions at $M_Z$,
the physical masses of the bottom quark and $\tau$ lepton,
and the value of the strong coupling at $M_Z$.  Therefore, solutions
for $B_0$, $\mu_0$, $y_\tau(M_X)$, $\alpha_3(M_X)$, and $M_t$ are
found consistent with the RG, the above constraints, and specified values for
the free input parameters.  We take the value of the strong coupling at $M_Z$
to be $\alpha_3(M_Z)=.118$.  The corresponding value of the strong coupling
at $M_X$ is determined based on this low energy constraint.  The values of
$\alpha_1(M_X)$ and $\alpha_2(M_X)$ are set equal and fixed at $1/25.3$.
This constant value for $\alpha_{1,2}$ at $M_X$ never leads to more than
about $1\%$ and $3\%$ error in $\alpha_{em}$ and $\sin^2\theta_W$,
respectively.  The difference in $\alpha_3(M_X)$ and $\alpha_{1,2}(M_X)$
is at most $3\%$ and can be accommodated using GUT thresholds.

Not all input values for the free parameters will yield adequate solutions,
and the $4+1$ dimensional parameter space must be explored and restricted
using various criteria.  Cases are rejected based on the existence of
color/charge breaking vacua or a charged lightest supersymmetric particle
(LSP).
In arriving at the superpartner mass bounds, the fine tuning prescription,
Eq.~(2.3), is applied to all solutions found in a grid of approximately 2000
points bounded as follows:  $|A_0|\leq 400$ GeV, $0\leq m_0\leq 400$ GeV,
$|m_{1/2}|\leq 500$ GeV, $1\leq\tan\beta(M_X)\leq 15$, and
${\rm sign}(\mu)=\pm$.

\section{Analysis}
\indent

The essential, novel feature of the fine tuning measure $\gamma$ is to
evaluate the sensitivity, $c$, of a physical quantity relative to a
benchmark, $\bar{c}$.  We have derived a formula for this benchmark in
section three and in Ref.~[1].
This prescription for calculating $\bar{c}$ requires us to choose a
range of integration $(a_{-},a_{+})$.
We use two conditions to define a suitable range of integration.  First,
we integrate over the all values of $a$ where
$SU(3)\times SU(2)\times U(1)$ is broken to $SU(3)\times U(1)_{em}$.
The resulting limits on the range of integration generally come from
two conditions on the value of $M_Z$.  The minimum value of $M_Z$ cannot
be less than $0$, and its maximum value cannot exceed some upper bound,
often set by the requirement that sneutrino squared masses be positive.
Second, in our analysis we only consider points where we are able to
find a significantly large range of integration.
If the range of integration is not suitably large, we will fail in
our attempt to compare the sensitivity of $M_Z$, when $a$ is chosen
so that the value of $M_Z$ is $91.2$ GeV,
to the average  sensitivity.  Inspection of Eq. (2.6)
shows that in the  limit of vanishing $(a_{+}-a_{-})$, $\bar{c}$
approaches $c$, and $\gamma$ tends to one.   To eliminate spurious
calculations of $\gamma$,
we only consider cases were $\delta a = a_{+}-a_{-}$ exceeds
$a/4$ or $a/8$ for $M_Z(a)=91.2$ GeV.  We find that typically
this has the effect of removing points where $SU(3)$ only remains
unbroken as the result of a fine tuning.

Figures 2-9 display correlations between the superpartner
masses and fine tuning.
For each solution point, we computed
the fine tuning with respect to the common scalar mass,
the top quark Yukawa coupling, and the common gaugino mass. Then,
for each individual solution,
we define $\tilde{\gamma}$ as the largest of these fine tunings.
Many earlier studies of naturalness,
as well as employing measures of sensitivity instead of fine-tuning,
considered the naturalness of the  $Z$ mass with respect to
individual parameters separately.  This separation can lead to a
significant underestimate of fine tuning.
\footnote{In fact, the original bound of $c<10$ imposed by Barbieri
and Giudice can no longer be satisfied. A calculation of the
sensitivity of the $Z$ mass with respect to $m_0$, $m_{1/2}$,
$y_t$, and $g_3$ gives $\tilde{c} > 30$ (see Fig.~1).}
In particular, we have compared the lower envelopes defined by scatter
plots, and we find explicitly that,
if fine tuning is plotted as a function of a particular coupling or
mass, the envelope defined by $\tilde{\gamma}$ cannot in general
be constructed from the individual envelopes for $\gamma(m_0)$,
$\gamma(m_{1/2})$, and $\gamma(y_t)$.
\footnote{This is a reflection of the fact that because the
$Z$ mass depends on several parameters, even if another variable
is fixed, it is easy to find solutions where the $Z$ boson's
dependence on an isolated fundamental parameter is relatively
insensitive.}
Figures 2-9 display the fine tuning measure $\tilde{\gamma}$ plotted
against selected superpartner masses.
The individual points shown in these figures correspond to
the grid of approximately 2000 points discussed in section three.
We caution the reader that the density of these points is
{\it not} an indication of how likely particular values of the
superpartner masses or $\gamma$ are.  This is because the grid we
have used is not completely uniform, and more importantly because
the minimization conditions (3.4-5) have been used to determine the
values of $B_0$ and $\mu_0$.  The dashed and dotted curves in these
figures show the minimum fine tuning necessary for a
particular value of the superpartner mass.  The likelihood or
naturalness of a particular value of a superpartner mass scales
like $1/\gamma$.

Figure 1 contrasts the sensitivity parameter $c$ with our measure
of fine tuning $\gamma$.   We see that currently viable solutions
depend on at least one fundamental parameter in a fairly
sensitive manner, however the fine tuning curve,
${\tilde\gamma}$, shows that this sensitivity is not always unusual.

Figures 2a-2b display the correlation between the gluino mass and the
fine tuning parameters $\tilde{\gamma}_1$ and $\tilde{\gamma}_2$.
This plot and, unless otherwise noted, the following plots
are constructed from solution points consistent
with the current LEP limits on superpartner masses \cite{LEP-I}.
We have taken the limits  on the sneutrino and the
charged superpartner masses to be $M_Z /2$, and the
lower limit on the light Higgs mass as $60$ GeV.
If no superpartner masses lie below these limits the most natural
value of the gluino mass is about $260$ GeV,  above the published
CDF limit of $141$ GeV and also above the recently reported
limit from D$\O$~\cite{CDFD0}.  For  potential, future search
limits at the Tevatron see for example Ref. [12].
If we require that fine tunings are at most a part in ten,
the gluino mass should lie below $\sim 600-800$ GeV, a value that
should be easily accessible at the LHC \cite{LHC}.

Figures 3a-b displays the correlation between fine tuning and
the lightest squark mass of the first and second
generation.  The analogous plots for the top squark mass are
shown in Figs.~4a-b. The most natural value of the stop mass
is around $220$ GeV and for the lightest of the remaining squarks
it is about $240$ GeV.  This is close to the preliminary mass
limit reported by D$\O$ at Glasgow \cite{CDFD0}.
If we require that fine tunings are at most a part in ten, the
stop mass should lie below $\sim 500-600$ GeV and the lightest of
the remaining squark masses should lie below $\sim 600-800$ GeV.

Figures 5a-b display the correlation between fine tuning and
the lightest chargino mass.  This plot displays solution points
consistent with the LEP derived constraints on superpartner masses
with the exception of the chargino mass.
The most natural value of the lightest chargino mass,
corresponding to the smallest $\tilde{\gamma}$, is around $50$ GeV.
Note that a significant region of the most natural solutions
lie within the physics reach of LEP II, which should be able to search
for charged particles up to the kinematic limit ~\cite{FS}.
The lightest chargino mass should not exceed $\sim 200-300$ GeV if
$\tilde{\gamma}<10$.

Figures 6a-b display the correlation between fine tuning and
the mass of the lightest superpartner.
The most natural value of the LSP mass appears to be around $42$ GeV,
and the theoretically favored values of the LSP mass are concentrated
below $70$ GeV.  The LSP can not be heavier than $150$ GeV if
${\tilde\gamma}<10$.  This bound provides a more stringent limit than
bounds set by the requirement that the LSP not over-close the
universe.

Figures 7a-b summarizes the mass predictions for all the superpartners.
The upper and lower ends of the bars correspond to ${\tilde\gamma}<10$
and the current experimental limits, respectively.  The diamond point
represents the ${\tilde\gamma}<5$ mass limit, and the square represents
the most natural value for the respective sparticle mass.

Finally, for completeness we display the correlation between
the lower bound on fine tuning and the fundamental
parameters $m_0$ and $|\mu_0|$ in Figs. 8 and 9.

\section{Conclusions}
\indent
As the mass limits on superpartners increase, it becomes increasingly
difficult to accommodate a light weak scale naturally.  We have
presented a detailed study of the relationship between superpartner
masses and naturalness.
This analysis demonstrates that supersymmetry
can not accommodate the weak scale without significant fine tuning if
superpartner masses lie beyond the physics reach of the LHC.
In addition our analysis reveals that the most natural values
of these masses often lie well below $1$ TeV.  We note that our limits are
higher than those which would be obtained using conventional
sensitivity criteria, but they lie below the bounds found in common
folklore.  In light of our results, we
feel the potential for the discovery of physics beyond the standard
model before the LHC is promising.
However, this optimism should {\it not} be
interpreted as a guarantee that LEP II or the Tevatron will see
superpartners even in the case when supersymmetry
is relevant to electroweak symmetry breaking.  A more detailed
application of these naturalness measures to collider SUSY discovery
reaches is in progress\cite{AC3}.

\section*{Acknowledgments}
We would like to thank Howie Baer and Xeres Tata for providing
us with updates on collider search reaches,
and GA would like to thank the Institute for Theoretical
Physics in Santa Barbara and the Aspen Center for Physics for
their hospitality.
This work was supported in part by funds provided by the
U.S. Department of Energy (DOE) under cooperative agreement
DE-FC02-94ER40818 and by the Texas National Research Laboratory
Commission under grant RGFY932786.

\newpage
\begin{description}

\item[\it Figure 1:] Curves representing the lower envelope of
regions defined by \hfil\break
${\tilde c}={\rm max}\{ c(m_{1/2}),c(m_0),c(y_t),c(g_3)\}$ and \hfil\break
${\tilde\gamma}_{1,2}={\rm max}\{\gamma_{1,2}(m_{1/2}),\gamma_{1,2}c(m_0),
\gamma_{1,2}(y_t),\gamma_{1,2}(g_3)\}$
plotted as a function of $\tan\beta$.    The upper curve
represents the amount of sensitivity required by current
experimental superpartner limits, and the lower curves display
the amount of fine tuning.

\item[\it Figures 2a-b:] The fine tuning measures $\gamma_{1,2}$ as a
function of the gluino mass.

\item[\it Figures 3a-b:] The fine tuning measures $\gamma_{1,2}$ as a
function of the lightest squark mass of the first two generations.

\item[\it Figures 4a-b:] The fine tuning measures $\gamma_{1,2}$ as a
function of the lightest top squark mass.

\item[\it Figures 5a-b:] The fine tuning measures $\gamma_{1,2}$ as a
function of the lightest chargino mass.

\item[\it Figures 6a-b:] The fine tuning measures $\gamma_{1,2}$ as a
function of the lightest sparticle mass.

\item[\it Figures 7a-b:] Superpartner mass ranges.  The upper and lower
ends of the bars correspond to ${\tilde\gamma}<10$ and the current
experimental limits, respectively.  The diamond (square) represents
the limit ${\tilde\gamma}<5$ (the most natural value).

\item[\it Figures 8a-b:] The fine tuning measures $\gamma_{1,2}$ as a
function of the common scalar mass $m_0$.

\item[\it Figures 9a-b:] The fine tuning measures $\gamma_{1,2}$ as a
function of the mixing parameter $|\mu_0|$.

\end{description}

\begin{thebibliography}{99}
%
\bibitem{AC1} G.~W.~Anderson and D.~J.~Casta\~no,
Phys. Lett. B {\bf 347}, 300 (1995) hep-ph/9409419.
%
\bibitem{BG} R. Barbieri and G. F. Giudice, Nucl. Phys,
{\bf B306}, 63 (1988);
J. Ellis, K. Enqvist, D.V. Nanopoulos, and F. Zwirner,
Mod. Phys. Lett. {\bf A1}, 57 (1986).
%
\bibitem{Wilson} K. Wilson, as quoted by L. Susskind, Phys. Rev.
{\bf D20}, 2619 (1979); G.'t Hooft, in {\it Recent developments
in gauge theories}, ed by G. 't Hooft et al. (Plenum Press, New
York, 1980)p. 135.
%
\bibitem{mr}See for example S.~Martin and P.~Ramond, Phys. Rev. {\bf D48},
5365 (1993).
%
\bibitem{ir}L. Ib\'a\~nez and G.G. Ross, Phys. Lett {\bf 110B}, 215 (1982).
%
\bibitem{ikkt}K.~Inoue, A.~Kakuto, H.~Komatsu, and S.~Takeshita,
Prog. Theor. Phys. {\bf 68}, 927 (1982).
%
\bibitem{agpw} L.~Alvarez-Gaum\'e, M.~Claudson, and M.~B.~Wise,
Nuc. Phys. {\bf B207}, 96 (1982); L.~Alvarez-Gaum\'e, J.~Polchinski,
and M.~B.~Wise, Nuc. Phys. {\bf B221}, 495 (1983).
%
\bibitem{ehnt}J.~Ellis, J.~S.~Hagelin, D.~V.~Nanopoulos, and K.~Tamvakis,
Phys. Lett. {\bf 125B}, 275 (1983).
%
\bibitem{cpr}See for example D.~J.~Casta\~no, E.~J.~Piard, and P.~Ramond,
Phys. Rev. {\bf D49}, 4882 (1994).
%
\bibitem{LEP-I}D. Decamp {\it et. al.} (ALEPH Collaboration),
Phys. Rep. {\bf 216}, 253 (1992); P. Abreu {\it et. al.}
(DELPHI Collaboration), Phys. Lett. {\bf B247}, 157 (1990);
O. Adriani {\it et. al.} (L3 Collaboration), Phys. Rep
{\bf 236}, 1 (1993);
M. akrawy {\it et. al.}
(OPAL Collaboration), Phys. Lett. {\bf B240}, 261 (1990);
G. Giacomelli and P. Giacomelli, Riv. Nuovo Cim. {\bf 16}, 1 (1993).
%
\bibitem{CDFD0}S. Hagopian FERMILAB-CONF-94-331-E, Oct 1994. 4pp.,
hep-ex/9410003; F. Abe {\it et. al.} (CDF Collaboration), Phys. Rev.
Lett. {\bf 69}, 3439 (1992); Phys. Rev{\bf D45}, 3921 (1992).
%
\bibitem{gluino}T. Kamon, J. Lopez, P. McIntyre
and J. White, CTP-TAMU-19/94 (1994); H. Baer, C. Kao and X. Tata,
Phys. Rev. {\bf D48}, R2978 (1993).
%
\bibitem{LHC}Preport of the Supersymmetry Working Group
(C. Albarjar {\it et. al.}) Proc of ECFA-LHC Workshop
(CERN 90-10), Vol II, 606-683 (1990).
%
\bibitem{FS} See for example J. Feng, and M. Strassler,
hep-ph/9408359 SLAC-PUB-6497, RU-94-67; H. Baer, C. Kao, and X. Tata,
Phys. Rev. {\bf d48}, 5175 (1993); M. Chen, C. Dionisi, M. Martinez
and X. Tata, Phys. Rep. {\bf 159}, 201 (1988).
%
\bibitem{AC3} G.~W.~Anderson and D.~J.~Casta\~no,
MIT-CTP-2464 hep-ph/9509212.
%
\end{thebibliography}
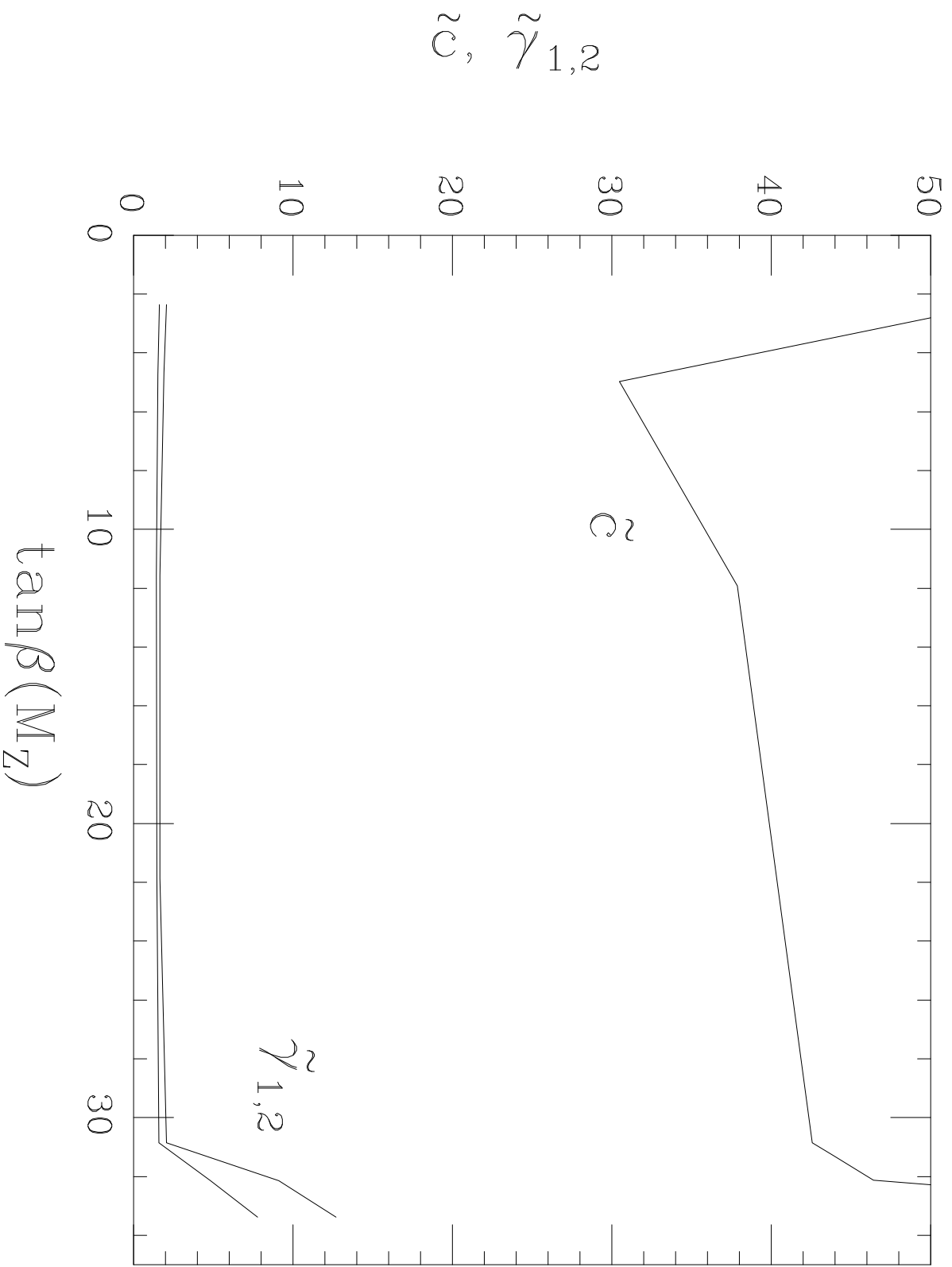
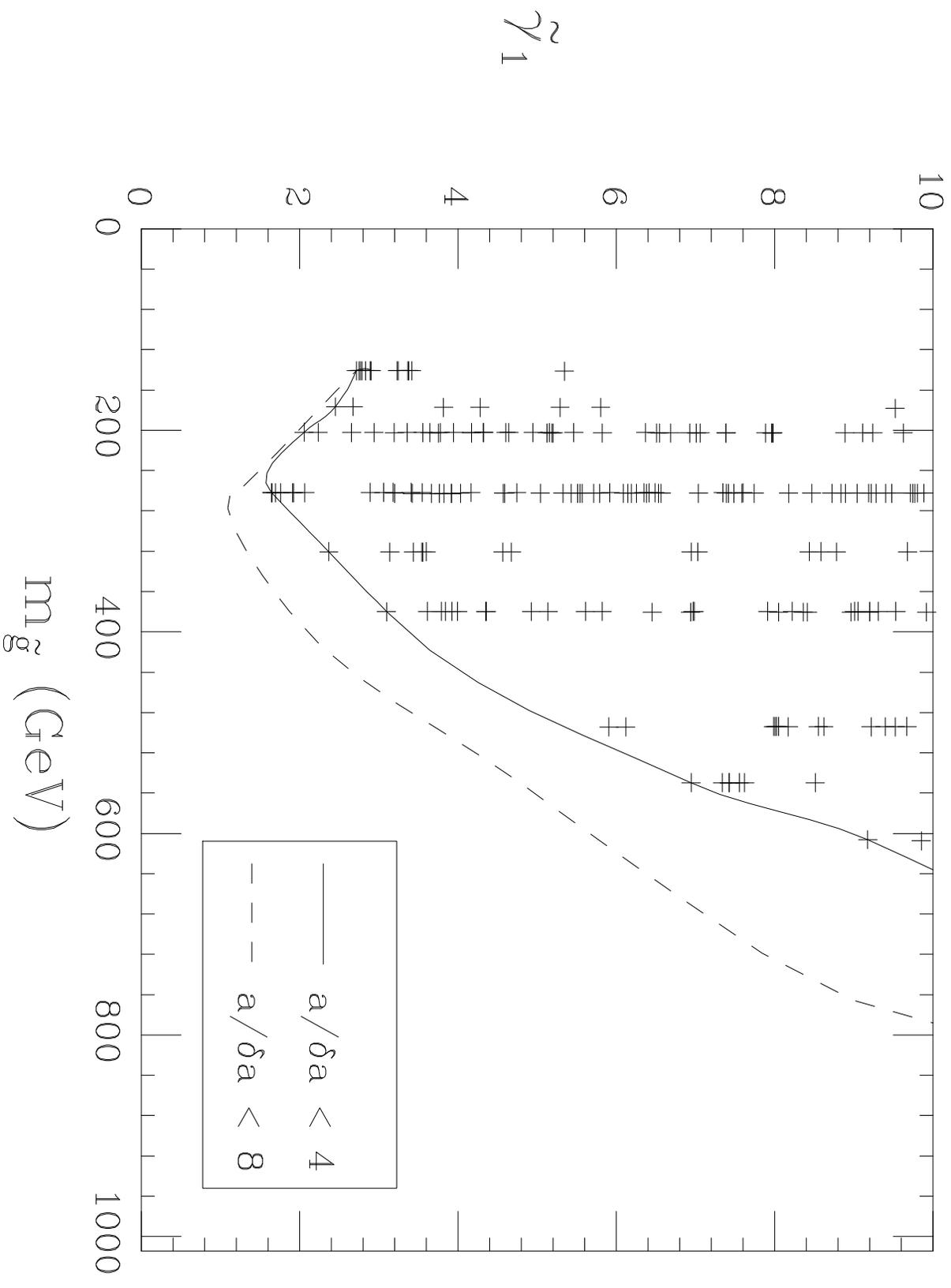
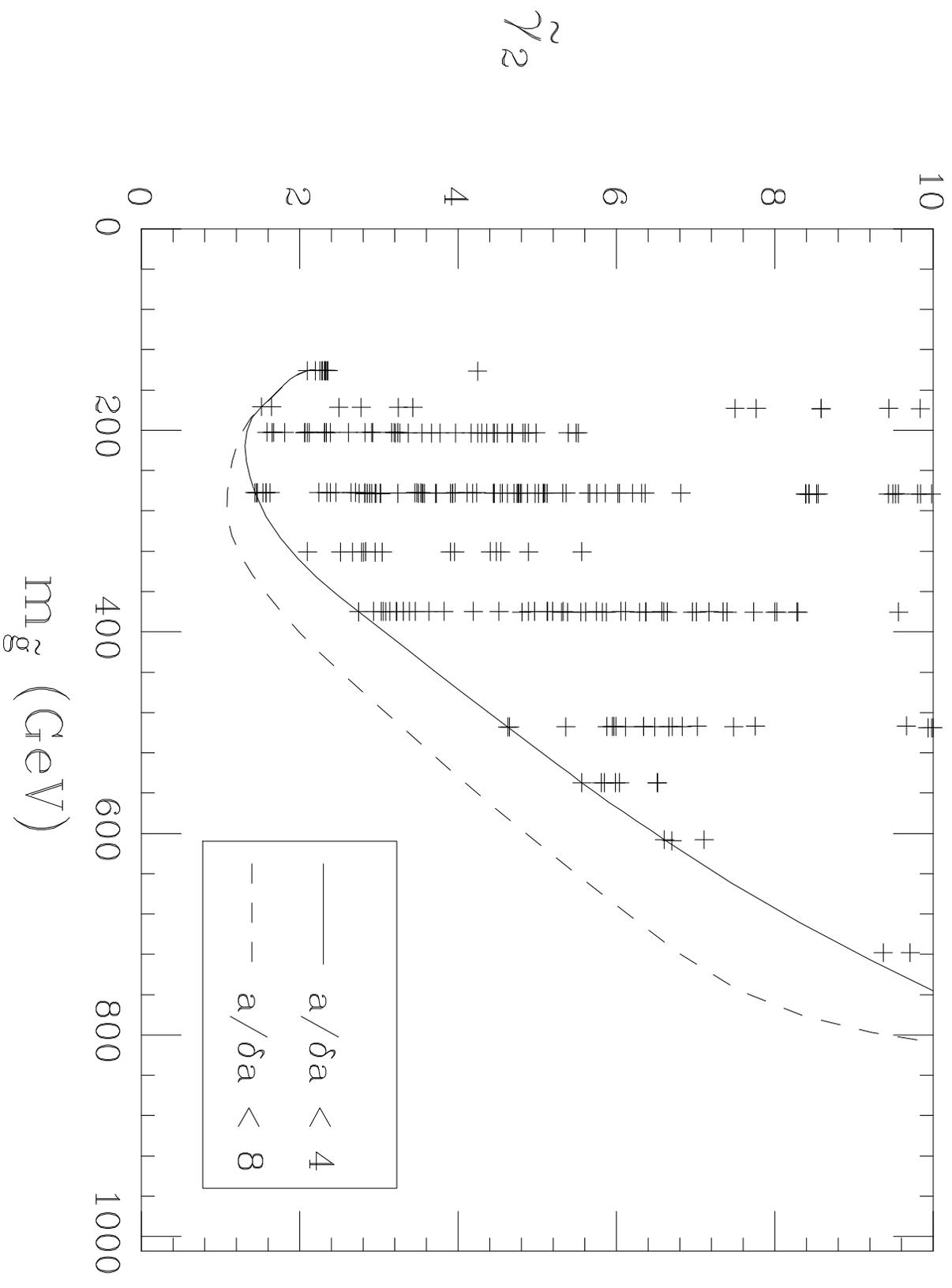
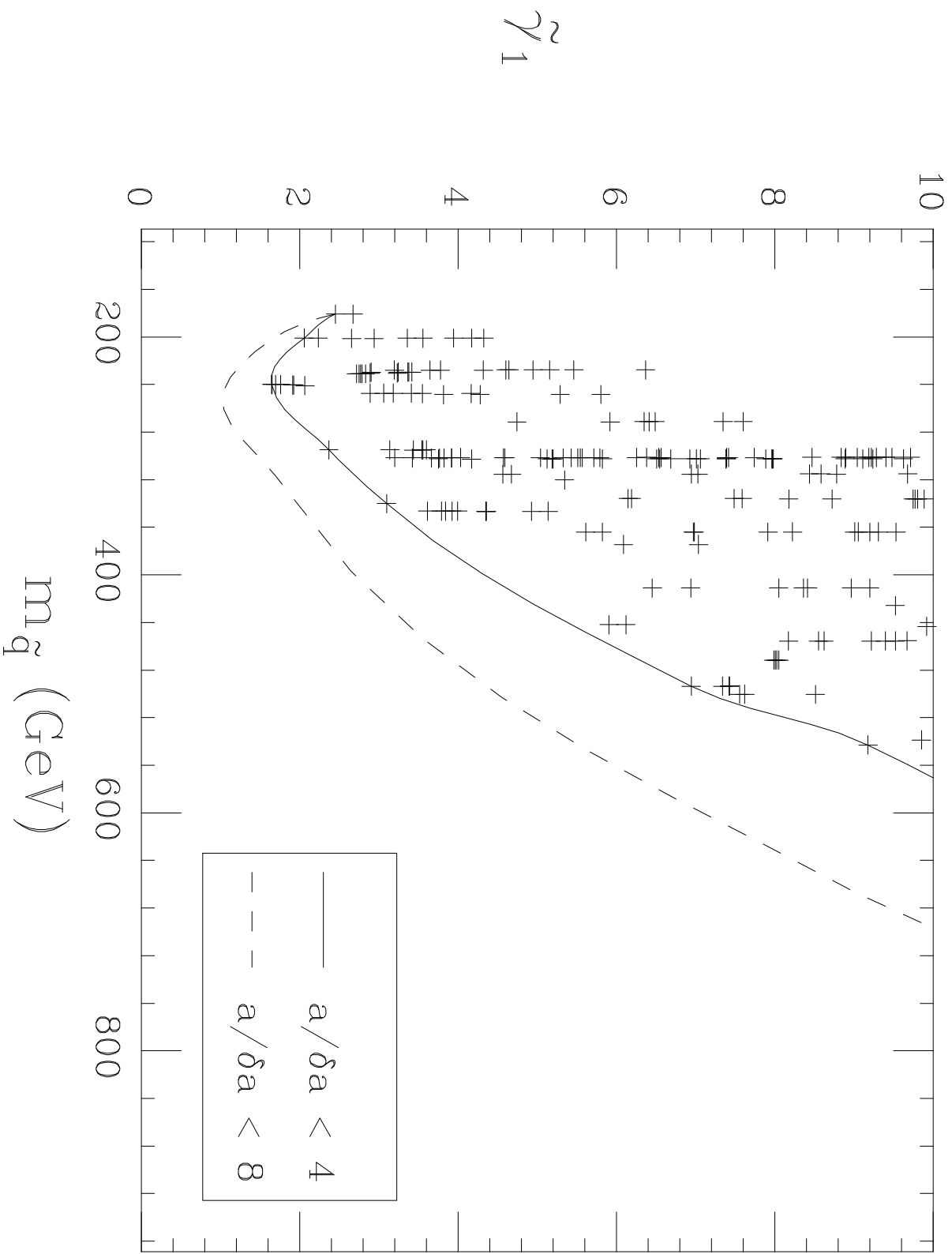
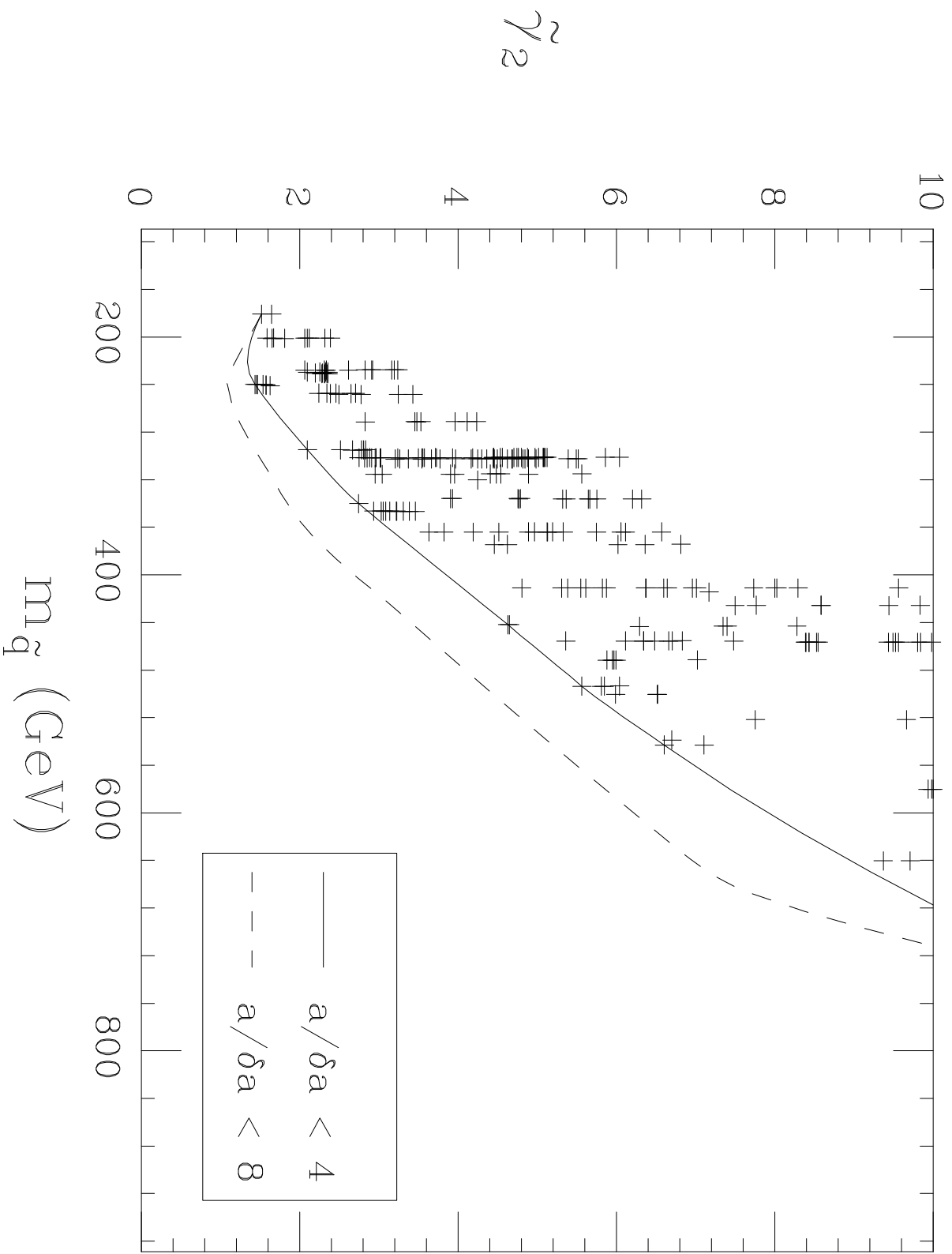
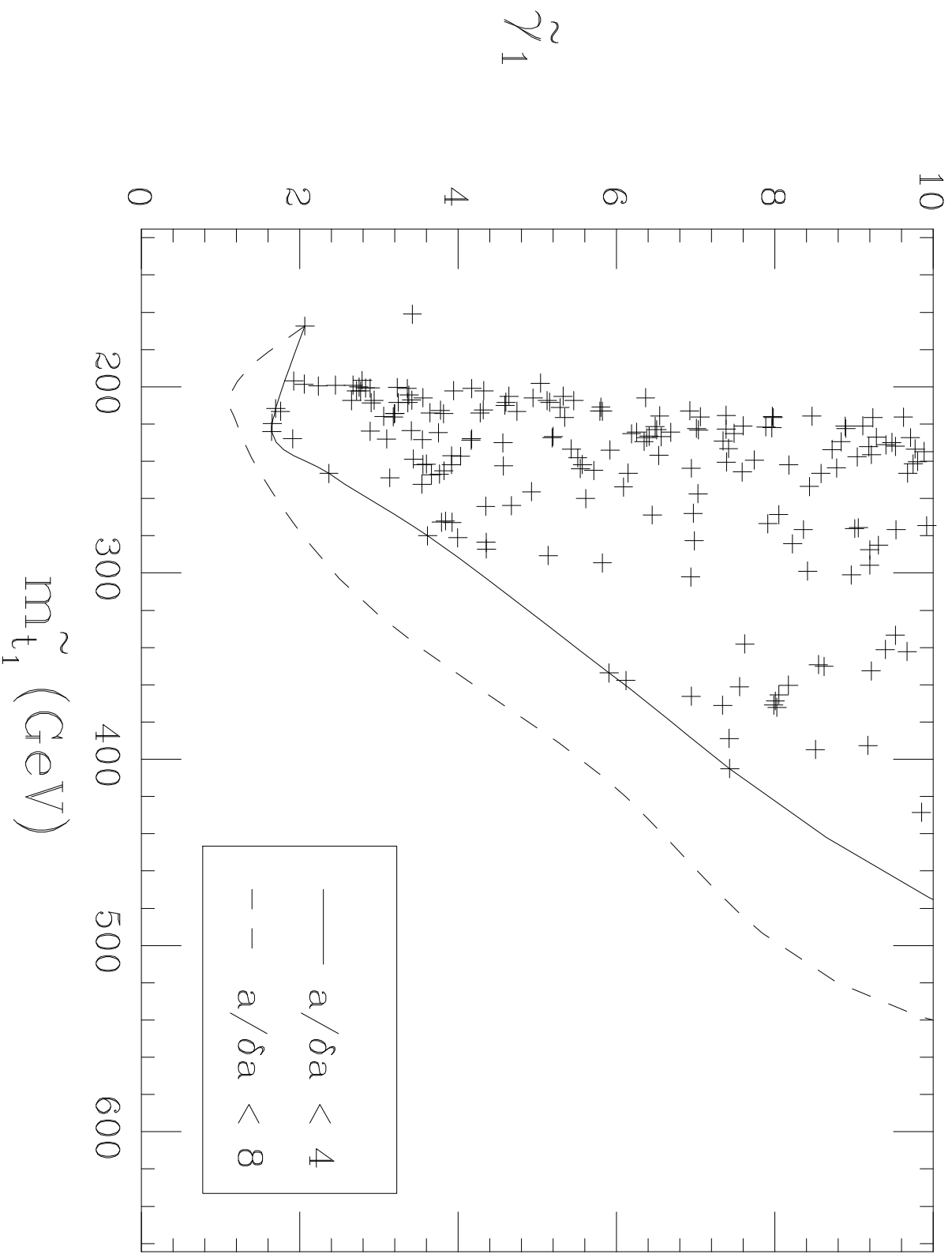
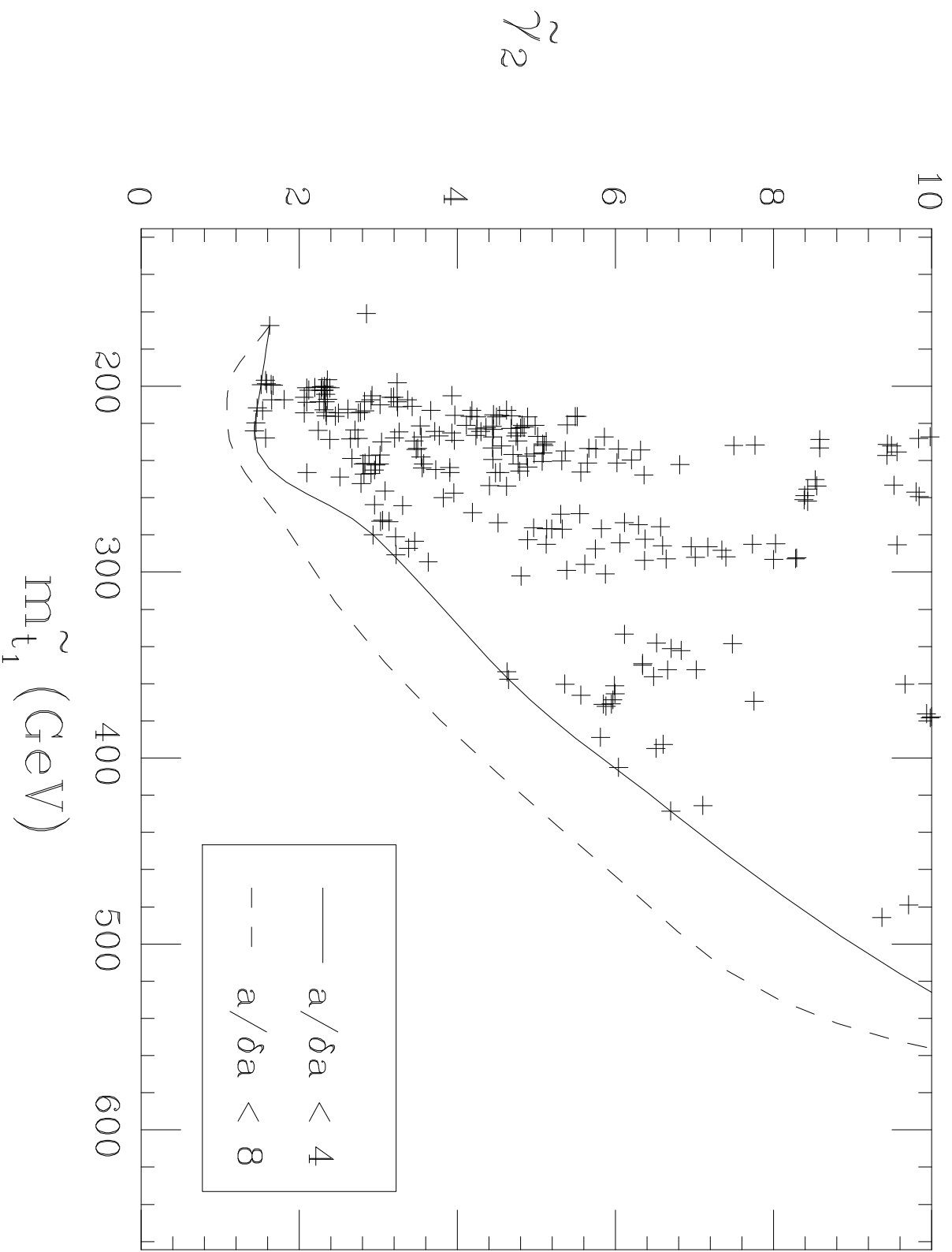
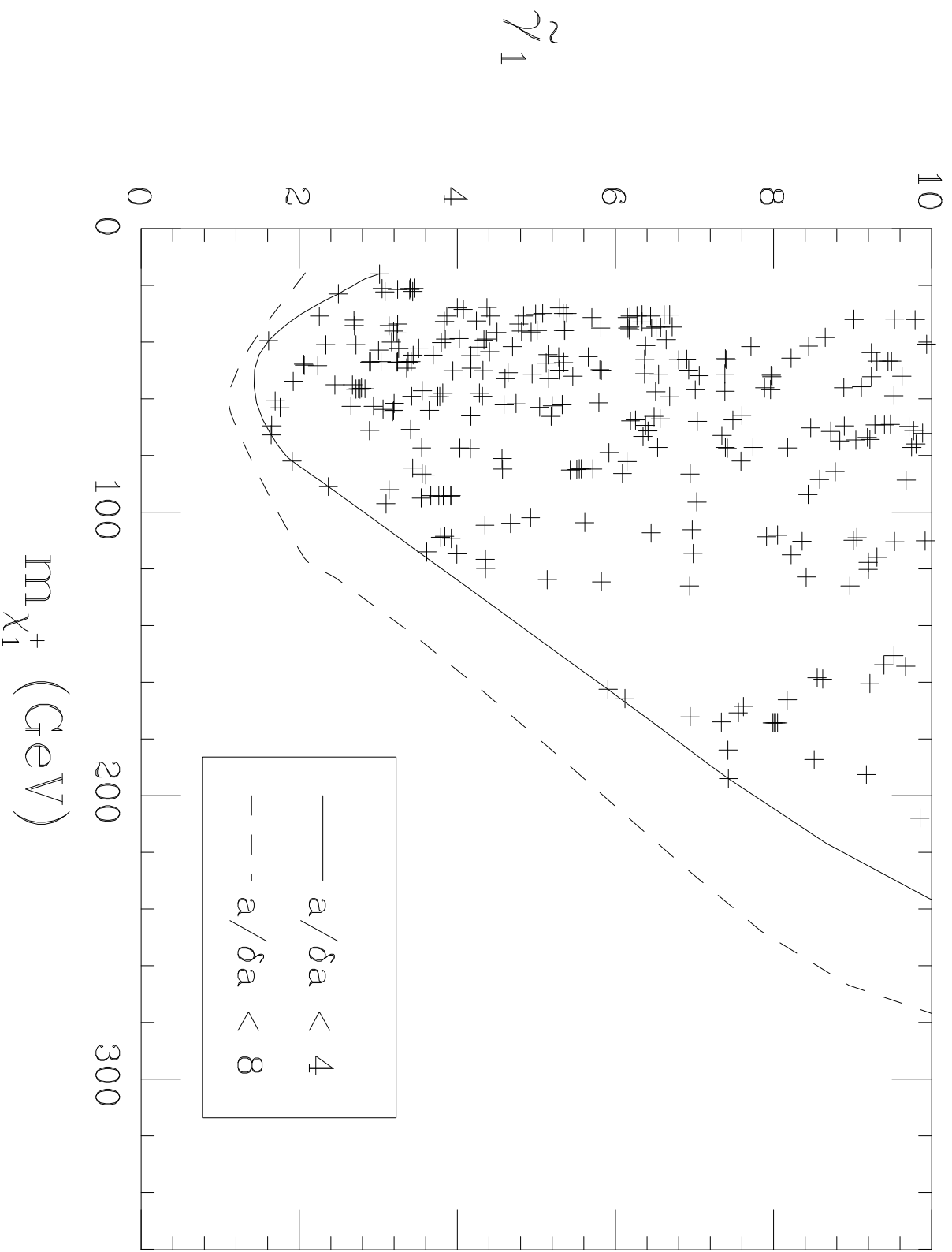
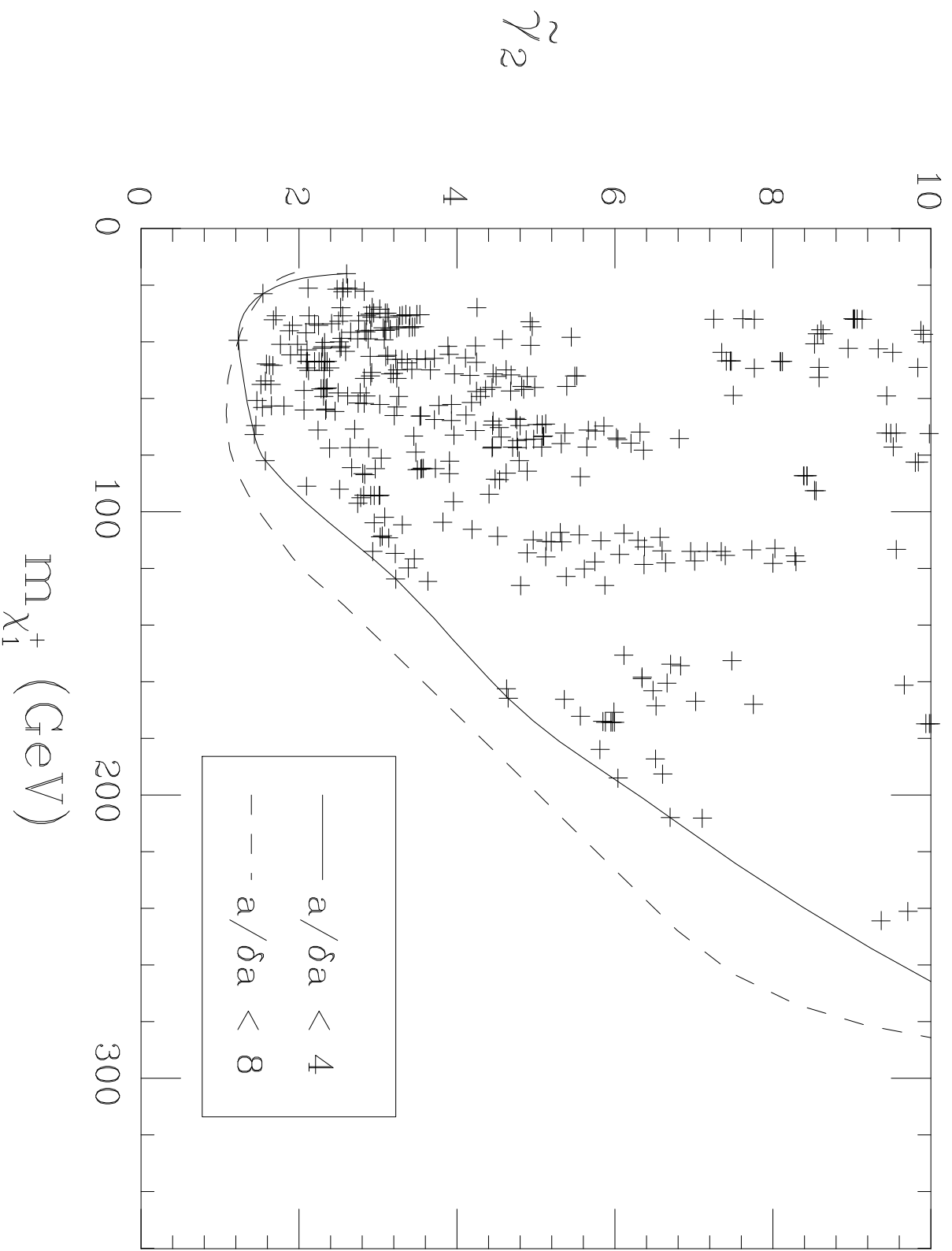
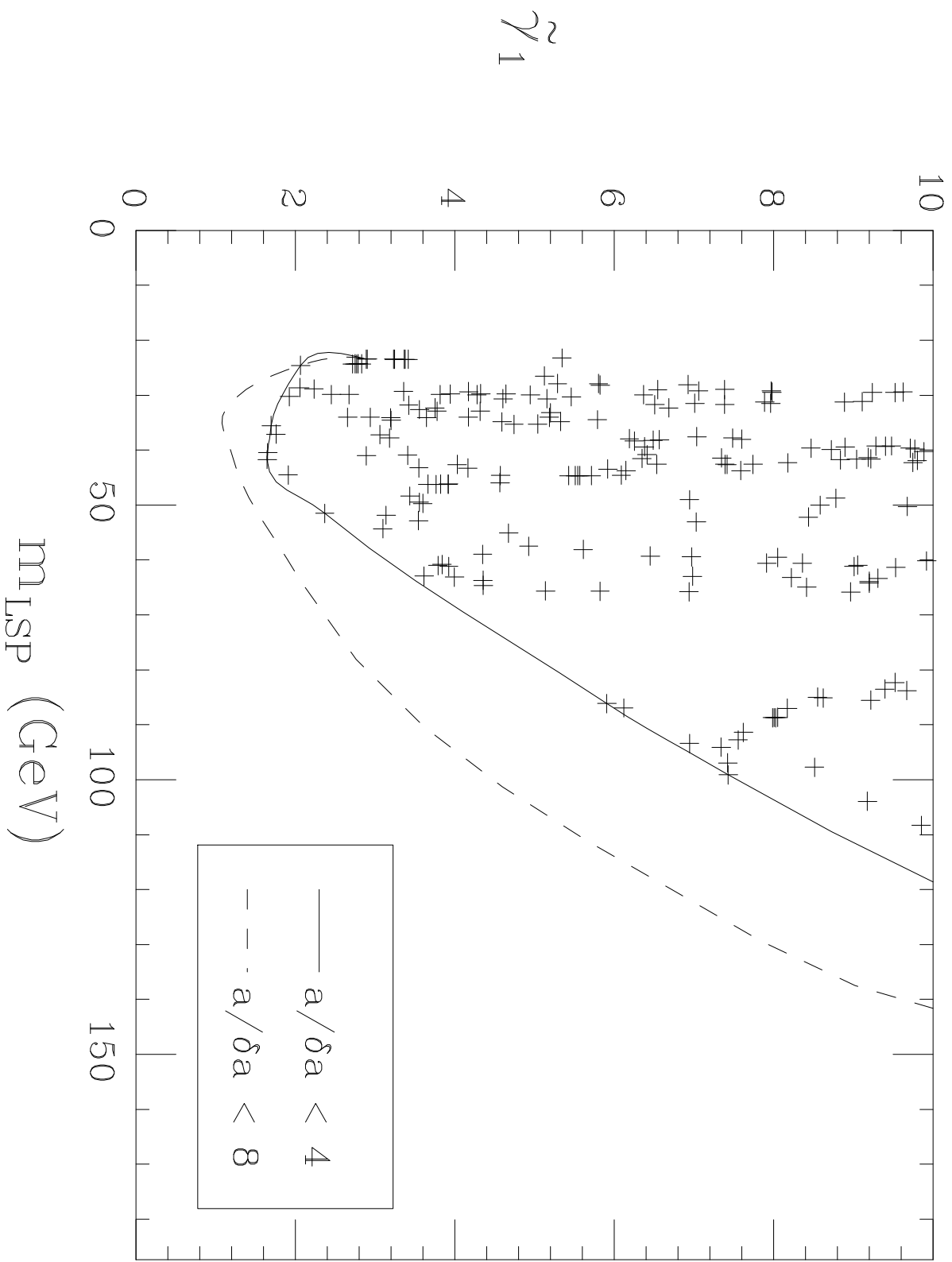
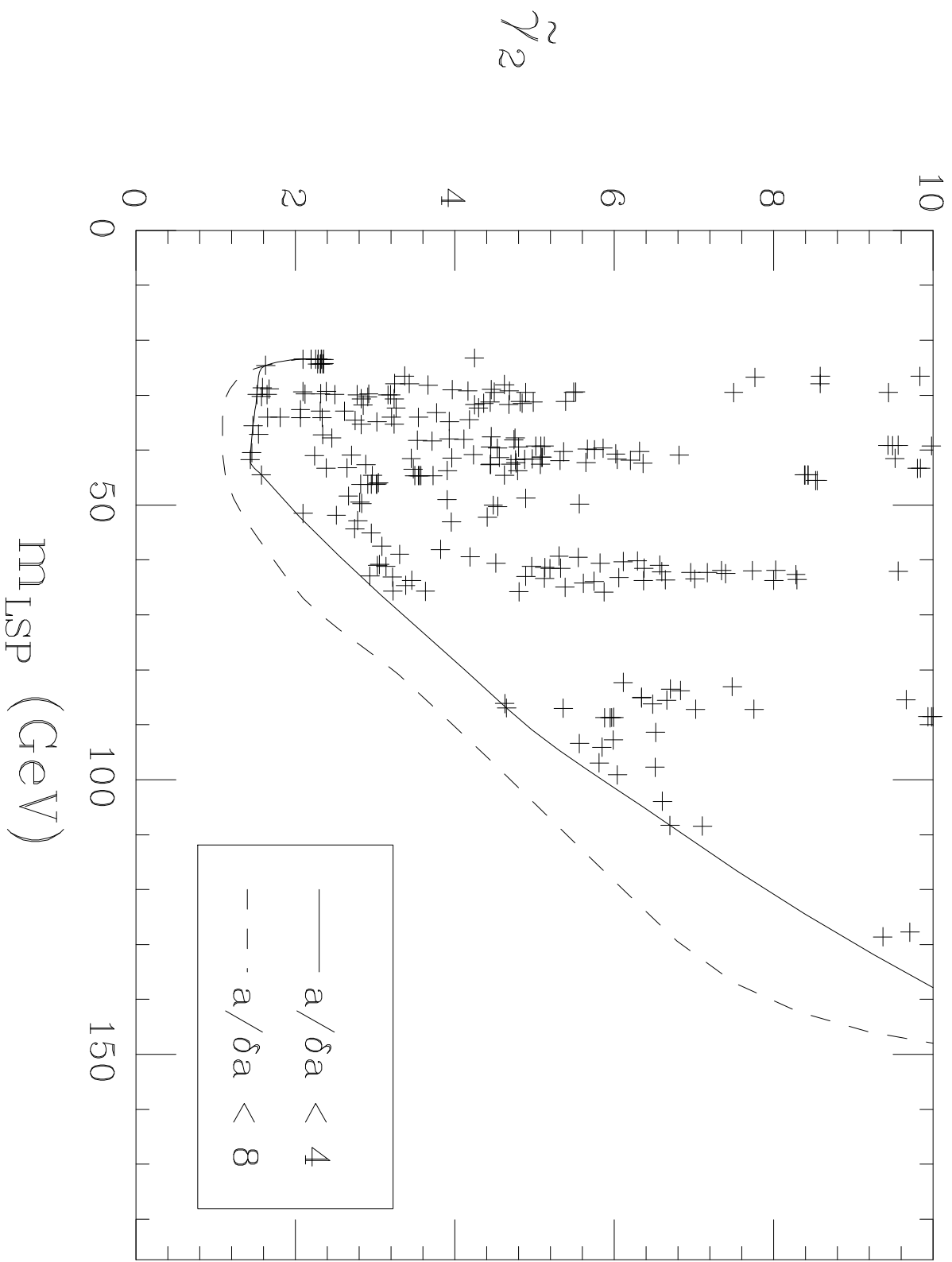
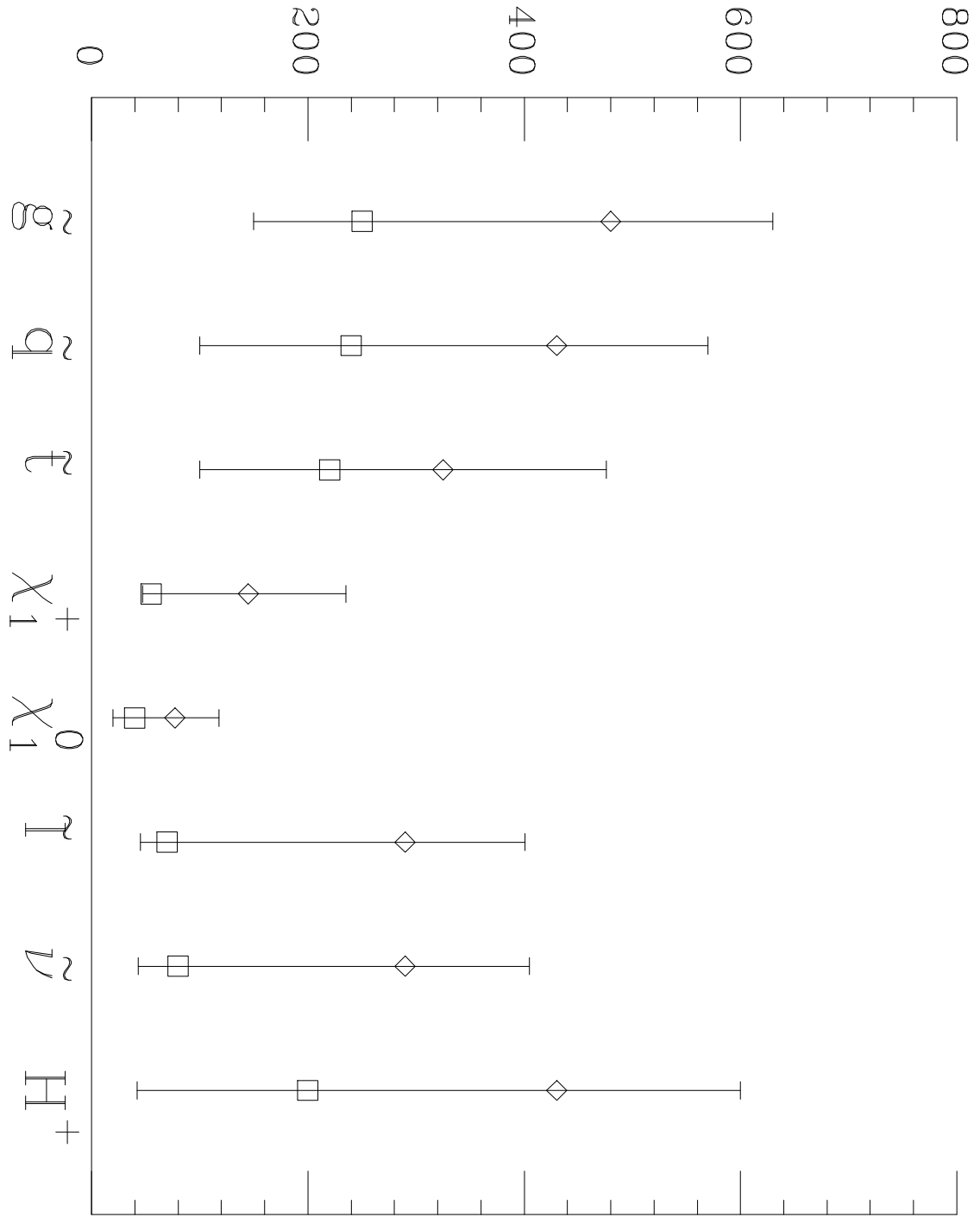
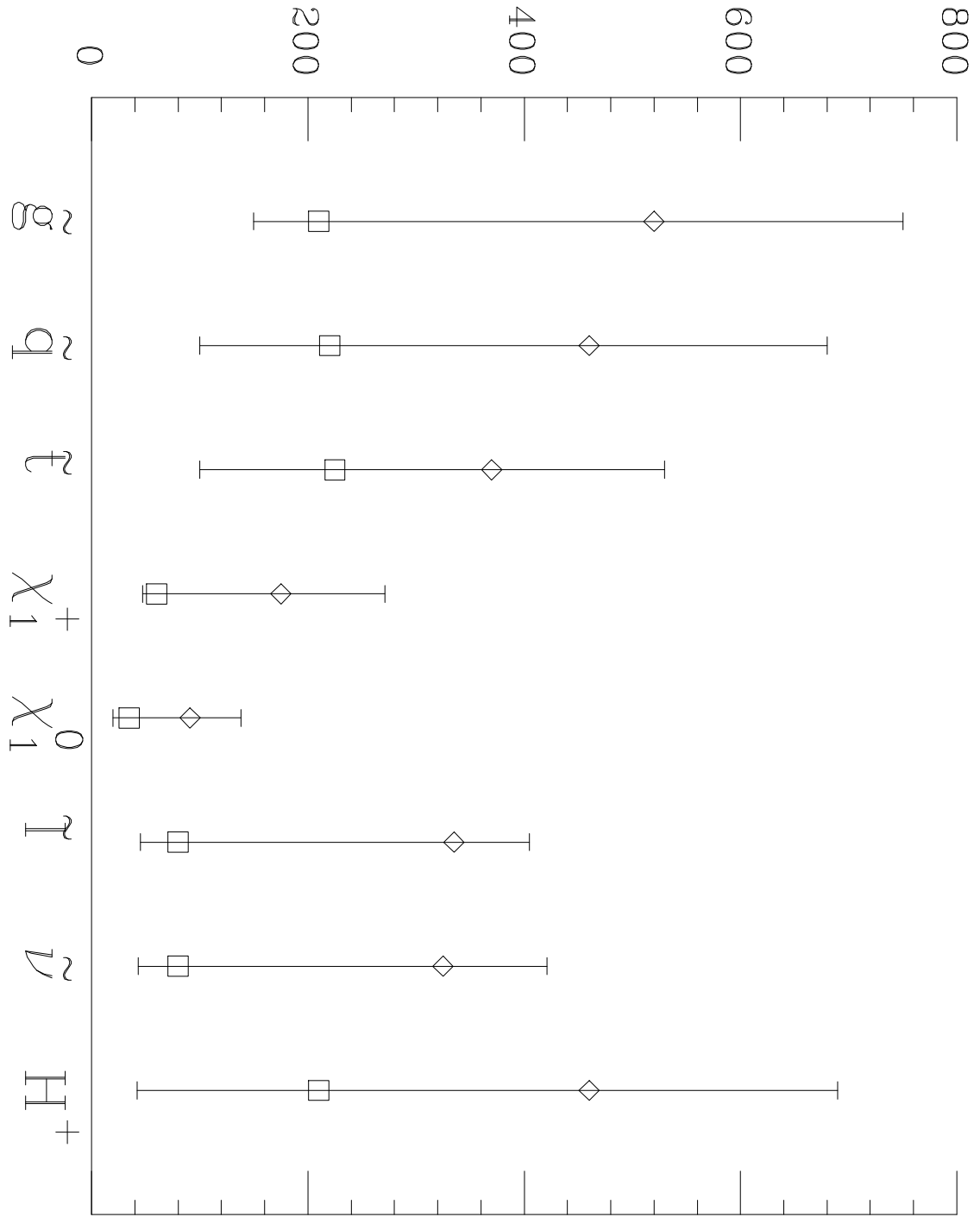
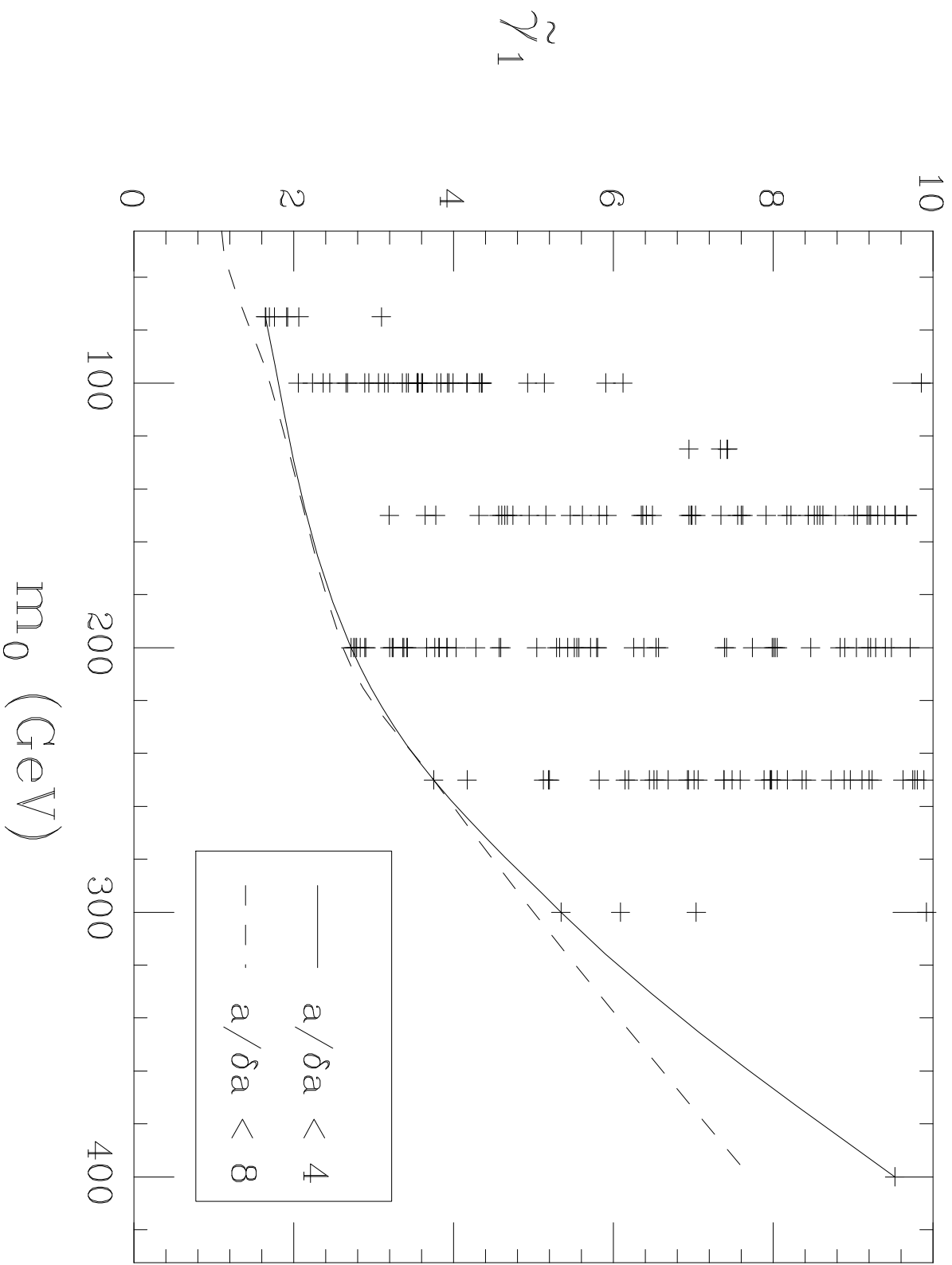
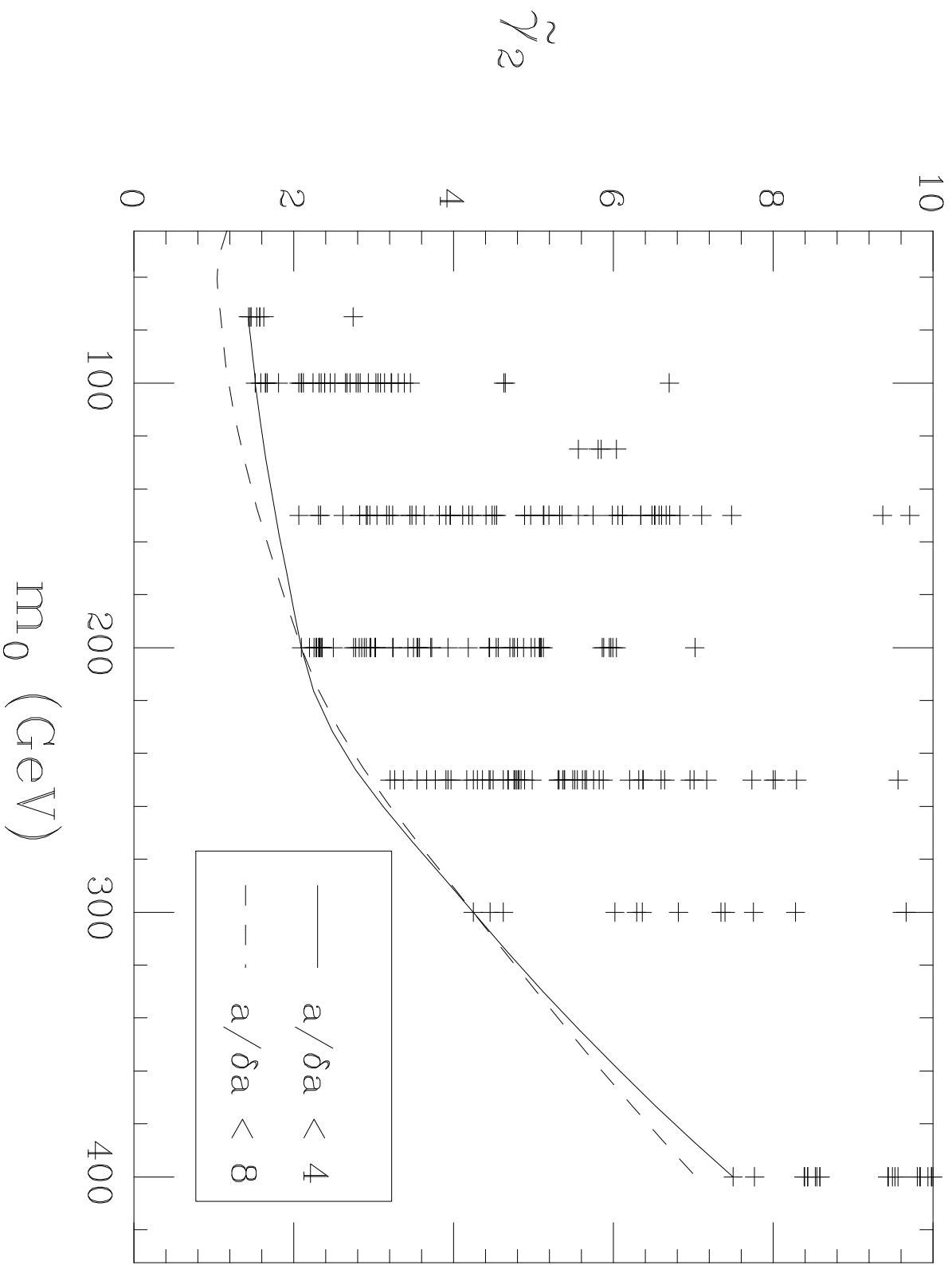
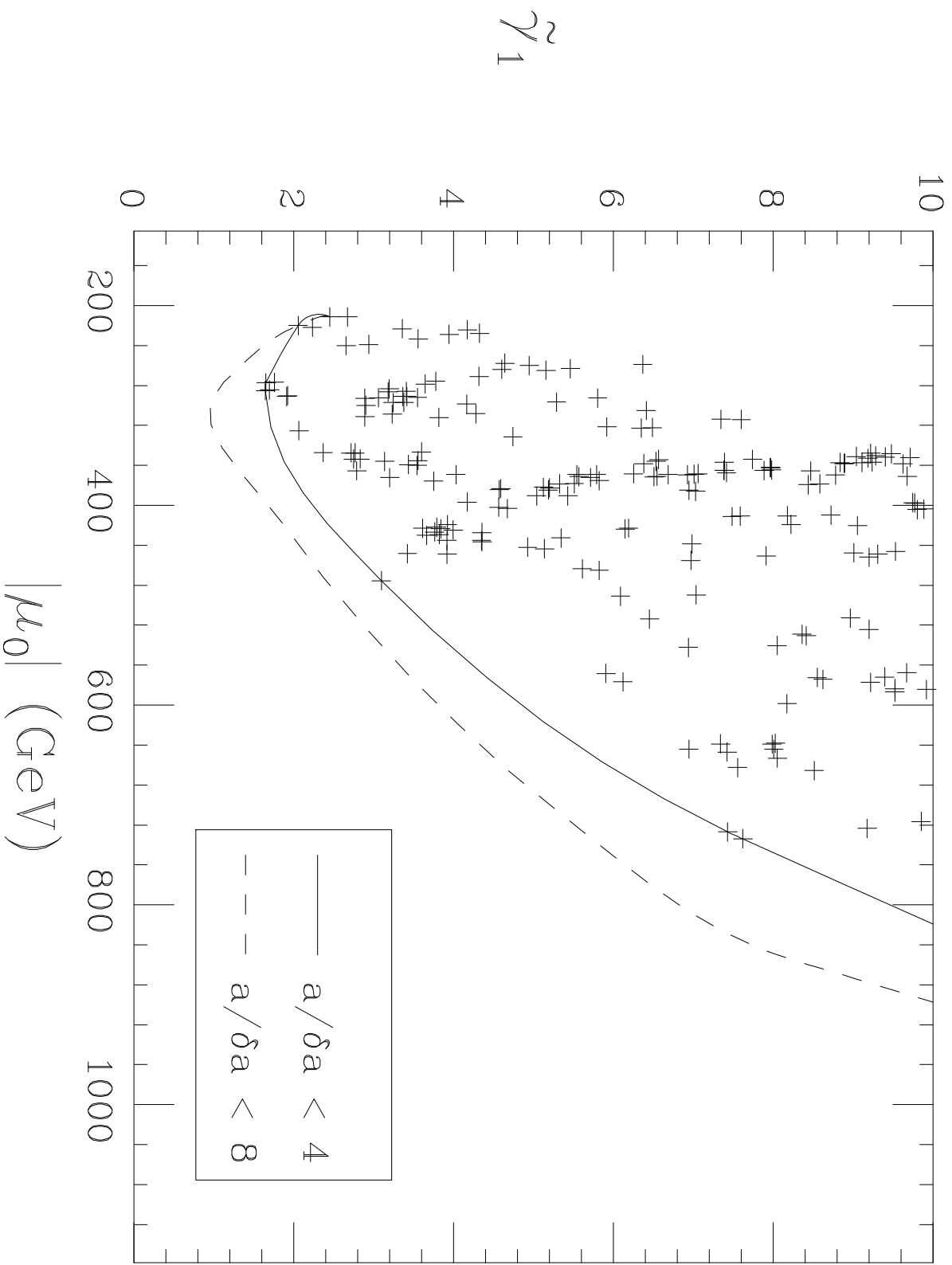
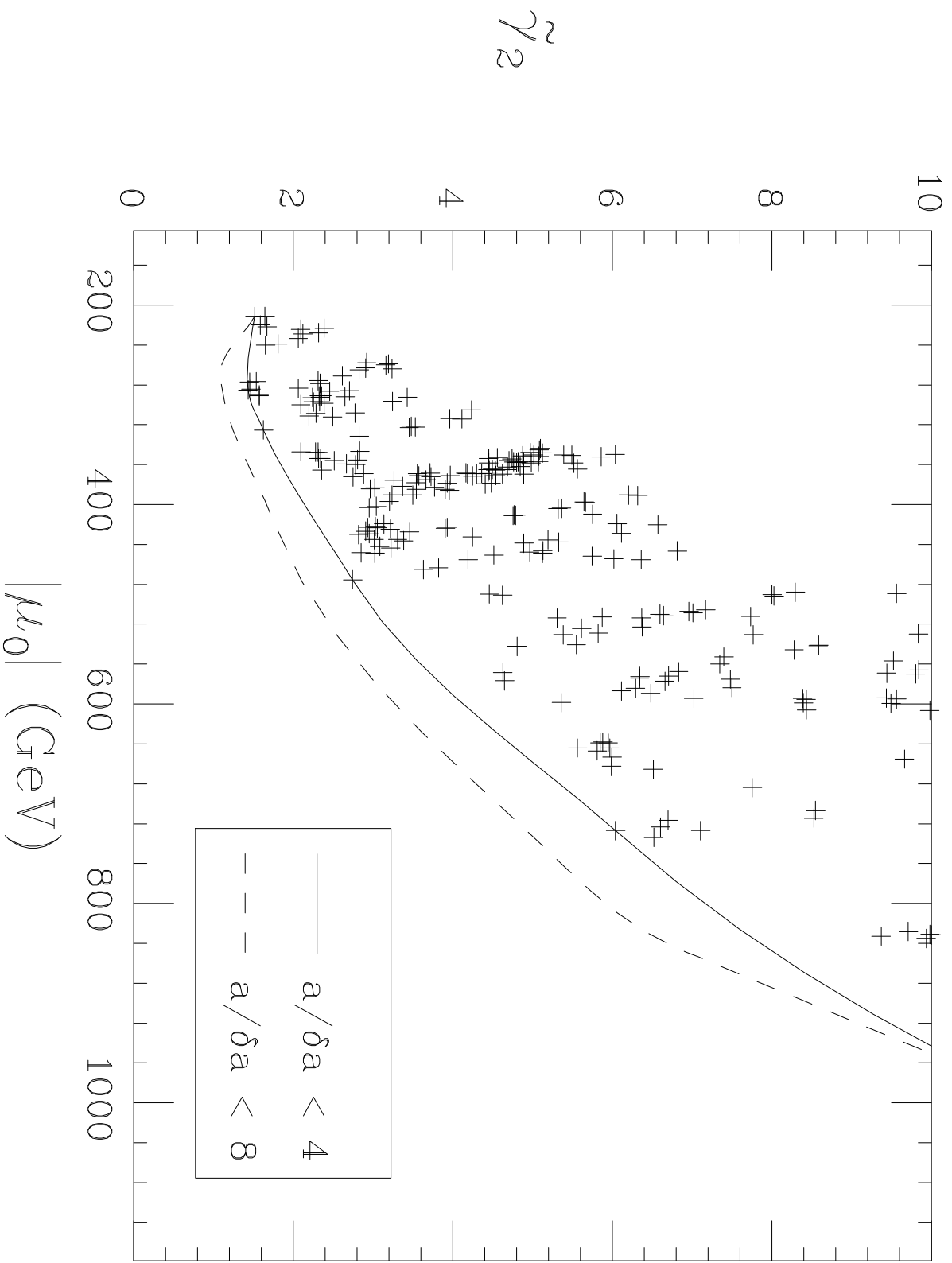
\end{document}